# Habitable worlds with no signs of life

Charles S Cockell, UK Centre for Astrobiology, University of Edinburgh, Edinburgh, EH10 4EP

Tel: 0131 650 2961; Email: c.s.cockell@ed.ac.uk

'Most habitable worlds in the cosmos will have no remotely detectable signs of life' is proposed as a biological hypothesis to be tested in studies of exoplanets. Habitable planets could be discovered elsewhere in the Universe, yet there are many hypothetical scenarios whereby the search for life on them could yield negative results. Scenarios for habitable worlds with no remotely detectable signatures of life include: planets that are habitable, but have no biosphere (Uninhabited Habitable Worlds); planets with life, but lacking any detectable surface signatures of that life (laboratory examples are provided) and planets with life, where the concentration of atmospheric gases produced or removed by biota are impossible to disentangle from abiotic processes because of the lack of detailed knowledge of planetary conditions (the 'problem of exoplanet thermodynamic uncertainty'). A rejection of the hypothesis would require that the origin of life usually occurs on habitable planets, that spectrally detectable pigments and/or metabolisms that produce unequivocal biosignature gases (e.g. oxygenic photosynthesis) usually evolve and that the organisms that harbour them usually achieve a sufficient biomass to produce biosignatures detectable to alien astronomers.

# 1. Introduction

One of the most remarkable future possibilities in biological sciences is the potential discovery of a second evolutionary experiment. Places to search for such a phenomenon are planets in our own Solar System and planets orbiting distant stars (exoplanets).

We make the assumption that planets most likely to exhibit biosignatures of life are those with the requisite environmental conditions to support the type of life that we are familiar with on the Earth [1]. These include: a temperature and pressure range conducive to the existence of liquid water as a solvent; the stability of complex carbon molecules required to build organisms; a diversity of elements required to carry out biochemical reactions, build molecules and act as catalysts (C, H, N,O, P and S as well as a range of metals such as Fe); the presence of energy sources, for example sunlight on the planet's surface or redox reactions in its crust and environments where these requirements can be provided over geological periods of time. Rocky planets within the habitable zone (the zone around a star where liquid water is stable at the planet's surface [2,3]) are one class of objects most likely to harbour another evolutionary experiment. Other planetary bodies can also host sources of liquid water, such as the interior of icy moons orbiting a gas giant far outside the classical habitable zone [4,5]. However, they are more difficult to assay for life due to their small size and the difficulty of detecting life using surface features or exhaled gases.

In recent years, efforts to detect rocky planets have yielded a diversity of candidates, some in the habitable zone [6,7], raising the possibility of investigating them for the existence of life. The two methods by which this is to be done are to search for surface signatures of life in the reflectance spectra of planets and to search for gases within the atmospheric spectra of planets that are produced by biological metabolisms. In the former case, the method relies on surface-dwelling life possessing molecules that absorb in specific regions of the spectrum to yield spectral features not expected from known abiotic materials [8-10]. In the latter case, the method relies on identifying gases that are thought to be only derivable from biological processes, for example high concentrations of oxygen [11-20].

If the search for other evolutionary experiments is to be undertaken consistent with the scientific method, astrobiologists must construct experimentally testable and falsifiable hypotheses that can be investigated by astronomers. In this paper, I discuss such a proposed hypothesis: 'Most habitable worlds in the cosmos will have no remotely detectable signs of life'. In sections 2-4, I identify scenarios that might lead to an acceptance of the hypothesis. In section 5, I identify the conditions required to reject the hypothesis.

## 2. Uninhabited habitable worlds

One potential class of rocky exoplanets that are conducive to life, but have no remotely detectable signatures of life are habitable, but lifeless worlds ('Uninhabited Habitable Worlds') [21].

The physico-chemical conditions necessary for the emergence of life are currently not sufficiently constrained to list them, or to know whether the emergence of life is inevitable on a planet that has habitable conditions [22].

Figure 1 shows four hypothetical scenarios that could lead to an Uninhabited Habitable World. These four scenarios are:

1) Planets that have habitable conditions, yet are too young for life to have originated. For such planets to exist, it would have to be the case that the appearance of life does not necessarily occur contemporaneously with the appearance of habitable conditions on a planet (i.e. once stable bodies of liquid water have formed). There must be a delay between the two events. Reliable evidence in the fossil record for the appearance of life ~3.47 Ga ago [23], yet the existence of potentially habitable conditions well before that, following late heavy bombardment [24], is potential evidence for such a delay on the Earth. However, it might also reflect the lack of preservation of the earliest life in the rock record or alternatively, an insufficient rock record.

2)      If the origin of life is rare because it requires an unusual concatenation of conditions that are found very infrequently on habitable planets, then many habitable planets might remain lifeless. For such worlds to exist, the origin of life must not always be inevitable when habitable conditions arise. As we are not fully aware upon what chemical contingencies the origin of life relies [22], we cannot quantitatively assess this possibility.

3)      Planets might lack appropriate conditions for an origin of life. For example, if the origin of life requires a land-based volcanic geothermal environment [25], then habitable exoplanets completely covered in ocean ('ocean worlds' [26,27]) might never yield an origin of life. As we remain ignorant to what diversity of environment(s) can initiate an origin of life, or whether it requires very specific planetary environments, this possibility cannot be assessed.

4)      A planet with unstable conditions might transiently possess habitable conditions at different points in its history (and during such a phase be observed by astronomers), but may lack the sufficient longevity of conditions to allow for an origin of life. For such worlds to exist, the same conditions as in (1) above must be met; an origin of life takes time and does not arise spontaneously when habitable conditions come into existence.

In summary, the existence of a class of exoplanets that are Uninhabited Habitable Worlds is conjectural. Our lack of knowledge about the inevitability of the origin of life, and the time it takes to occur on habitable planets makes it impossible to quantitatively predict the proportion of habitable worlds that are lifeless, if they exist at all. Nevertheless, if such planets exist, they would be one class of worlds that exhibit habitable conditions, but have no remotely detectable signs of life.

Even if the origin of life does generally occur on habitable worlds, there are ways in which habitable and inhabited planets could still exhibit no remotely detectable signs of life. These types of worlds are the focus of the following two sections.

**3.     Inhabited worlds that have no detectable surface biosignatures**

One method by which life could be detected is to examine the reflectance spectra from habitable worlds. The absorption of regions of the spectrum by life, particularly in the visible and infra-red range, can produce spectral features not expected from known abiotic materials. An example is the chlorophyll *a* 'red edge', which results in a red absorption and infra-red reflectance from terrestrial chlorophyll *a*-containing organisms [8, 28, 29]; a signature of sufficient strength to be detectable in Earthshine [30]. Earthshine measurements indicate that the Earth's red edge is feasibly detectable by alien observers, but is made very difficult due to its broadness and the interfering effects of cloud coverage [31, 32].

Although such a signature can be remotely detected on the Earth, what sort of habitable worlds would lack such signatures? Examples of these types of worlds are discussed below. Note that the examples are not mutually exclusive. A combination of the cases discussed in sections 3.1 and 3.2 could exist on a single planet, combining to compound problems of detection.

**3.1     Cryptic biota**

Much of the microscopic life on Earth lives beneath the surface and as such is undetectable in a planetary reflectance spectrum (for a full review of these modes of growth see [33]). Examples of this type of life include: deep subsurface biota inhabiting rocks and fracture zones [34, 35]; life inhabiting the subsurface of oceans [36-39]; life within soil particles [40]; and life within [41, 42] or under [43, 44] surface rocks.

On planets with extreme surface conditions, near-surface microorganisms will tend to grow within or under substrates where physical conditions are more clement [45]. Examples include organisms that inhabit the interior of salt deposits [46-49] and porous rocks [50-52], i.e. volcanic [53, 54] and impact-shocked rocks [55] (endoliths), as well as organisms that live on the underside of rocks [56, 57] (hypoliths).

One example of cryptic biota is photosynthetic organisms living within a few millimetres from the surface of a porous rock, as illustrated In Figure 2 with an impact-altered gneiss. The rock, collected from the Haughton impact structure in the Canadian High Arctic (see [55] for more details), has been heated and pressurised during an asteroid or comet impact event ~39 million years ago [58], resulting in a porous structure within which photosynthetic microorganisms (primarily cyanobacteria) can grow. Microorganisms are absent from the surface, which is desiccated and exposed to arctic winds, rendering it inhospitable to photosynthetic organisms. However, microorganisms inhabit the protected interior and form a narrow band of a few millimetres depth. The exact depth depends on the local rock structure; the lower limit of colonisation depth is set by light availability, with colonisation extinguished where light is below the minimum required for photosynthesis.

Using an Ocean Optics USB-2000 spectrometer (Ocean Optics, Manchester, UK) calibrated with a white reflectance PTFE (polytetrafluoroethylene) control surface (WS-1; Ocean Optics, Manchester, UK), a reflectance spectrum (400-800 nm) was obtained by positioning the light source (Ocean Optics HL-2000 Tungsten-Halogen fiber optic source) and probe over an exposed area of the rock interior within which organisms were growing. The spectrum yielded a 'red-edge', resulting from chlorophyll *a* within the organisms (Figure 2). However, the rock surface, directly above the same region of growth, only produced a relatively featureless spectrum of the gneiss rock (Figure 2). The final spectra were the mean of nine separate spectra acquired from three different locations within, or on the surface of, the endolith community. At each location the spectrum was the mean of ten spectral scans taken across the wavelength range. Standard deviation at any given wavelength in the spectra obtained was less than 5%.

The extent to which the red edge is distinguishable from spectra without a red edge can be quantified by calculating the mean reflectance across a given wavelength range before the red edge ($r_a$) and after the red edge ($r_b$). These ranges are typically taken to be 600-670 nm and 740-800 nm, respectively, and the magnitude of the vegetation red edge is defined as VRE = $(r_b-r_a)/r_a$ [59]. The VRE for the endolithic community studied here was 0.30. The same measurement for the gneiss surface that covers the community was 0.06 and similar values were obtained for control shocked gneiss surfaces (spectra not shown) from uncolonised rocks, illustrating the inability to detect cryptic biota.

Photosynthetic organisms were chosen to illustrate the concept of cryptic biota due to their distinct spectral features. They are an exceptional example since at sufficient biomass they are potentially detectable in atmospheric spectra as a consequence of their oxygen production [33]. However, the same principle applies to any organisms that possess a distinct spectral reflectance feature, yet are hidden beneath the planetary surface.

On planets where surface conditions are extreme and life is primarily cryptic, the lack of widespread surface life may be sufficient to render the integrated planetary disc spectrum devoid of any remotely detectable surface reflectance features.

### 3.2    Surface biosignatures at insufficient concentrations

Even if organisms do grow on a planet's surface, they will not necessarily be detectable by their effect on light reflected off the surface.

An obvious case is planets where an origin of life has occurred, however, life is at insufficient biomass to produce detectable surface signatures, because either a lack of time for organisms to widely and extensively colonise the planet, or a limitation of resources means that life never achieves sufficient biomass.

To explore the possibility that, even at high surface abundance, microbial communities can remain undetectable, artificial mixtures of organisms and quartz sand were created and their reflectance spectra investigated.

*Cupriavidus metallidurans*, inhabiting a diversity of environments, was chosen as a model organism. It is a metabolically versatile (capable of heterotrophy and chemolithotrophy), non-spore-forming, Gram-negative organism that inhabits heavy metal-contaminated sites [60, 61]. Representatives of its genus have been identified in volcanic ash and other harsh environments [61].

*C. metallidurans* was obtained from Deutsche Sammlung von Mikroorganismen und Zellkulturen (DSMZ) (No. 2839) and grown on minimal medium [62] agar plates. Pure colonies grown on plates (with a cell abundance of $\sim 10^{12}$ organisms), have some broad spectral features distinguishable from the quartz reflectance spectrum (Figure 3a), as determined using a spectrometer (as referred to in section 3.1).

Organisms were mixed with quartz sand (Dobbies Garden Centre, Edinburgh; particle diameters between 0.5 and 1 mm) at a concentration of $10^8$ organisms/cm$^3$, as determined by microscopy, and allowed to dry.

To quantify the degree to which the spectral features attributed to the organisms are distinguishable from the control sand, a similar calculation as that described for the red edge in section 3.1 was used. For *C. metallidurans*, the ranges 420 – 480 nm and 740-800 nm were chosen for $r_a$ and $r_b$, respectively as these wavelength ranges most clearly represented regions where the organisms had different spectral effects (Figure 3a, spectrum of pure colonies). The corresponding calculation of $(r_b-r_a)/r_a$, we might call the 'Pigment Spectral Effect' (PSE), thus the red edge is a particular type of Pigment Spectral Effect.

At a concentration of $10^8$ organisms/cm$^3$, none of the spectral features recorded in the pure colonies were observed (Table 1) and the spectrum was indistinguishable from the quartz sand spectrum (Figure 3a). A microbial concentration of $10^8$ organisms/cm$^3$ was chosen as it is representative of many fecund marine and soil environments [63, 64]. However, it is unlikely that such concentrations would exist across an entire planetary hemisphere; it would require the planet to be uniformly covered by a biomat of organisms. The data show that a '*Cupriavidus* world' (a world covered in *Cupriavidus*-like organisms) would likely be undetectable to alien observers.

The experiment was repeated using an organism with a distinctive spectral feature to determine whether this would significantly affect the detectability. *Deinococcus radiodurans* is an organism that stains Gram-positive (although its cell wall has Gram-negative characteristics). It belongs to a genus found in a wide diversity of environments, including deserts [65, 66]. It is resistant to high doses of ionizing radiation [67], which has been partly attributed to its effective protection and repair capabilities; a product of exposure to the desiccating environments in which it lives [68]. One of its characteristics is a high concentration of the carotenoid, deinoxanthin [69]. A function of carotenoids is to quench reactive oxygen species, which is beneficial for cells in desert environments exposed to solar radiation and transient desiccation. These molecules confer upon the organism a distinctive red colouration.

*D. radiodurans*, obtained from the DSMZ (No. 20539), was grown on tryptic soy agar plates. The spectrum of *D. radiodurans* biofilm grown on plates (Figure 3b) showed the presence of deinoxanthin, revealed as a 'green-edge'.

The organisms were mixed with quartz sand at a concentration of $10^8$ organisms/cm$^3$, as described for *C. metallidurans*. When the spectrum was measured, the characteristic green edge disappeared and the reflectance spectrum was indistinguishable from the quartz sand control (Figure 3b). This was quantitatively confirmed by calculating the Pigment Spectral Effect using the same wavelength ranges as described for *C. metallidurans* (Table 1). The data indicated that a '*Deinococcus* world' (a world covered in *Deinococcus*-like organisms) would likely be undetectable to alien observers.

Some organisms contain pigments, such as chlorophylls, that have evolved to absorb light very effectively. Figure 3c shows the same experiment repeated using *Chroococcidiopsis* (strain 029), a polyextremophilic cyanobacterium commonly found in deserts and in endolithic and hypolithic habitats as well as rock surfaces [70-72]. As with other oxygenic photosynthetic organisms that contain chlorophyll *a*, such as plants and algae, *Chroococcidiopsis* exhibits a characteristic red edge (Figure 3c).

At ecologically plausible concentrations of $10^6$ organisms/cm$^3$, the red edge was detectable in the quartz sand background (Figure 3b). This was quantitatively confirmed by calculating the vegetation red edge ($(r_b-r_a)/r_a$) as described previously in section 3.1 [59]. This observation may be accounted for by the large cell size (*Chroococcidiopsis* cells are ~3-4 μm in diameter, about 30-60 times the volume of *Cupriavidus* or *Deinococcus* cells, thus compensating for the lower cell numbers) and by the fact that chlorophylls have evolved specifically as efficient light-absorbing molecules for energy acquisition, making them effective in influencing spectral characteristics.

Spectral results for *Chroococcidiopsis* in sand are consistent with the data presented in Figure 2 showing the presence of a red edge in a natural endolithic community at a cell concentration ~$10^7$ organisms/cm$^3$. The data are also generally consistent with the detectability of the red edge in Earthshine [32, 73, 74] and the use of pigments for the remote sensing of oceanic primary productivity [75], and thus, demonstrate that some pigments are remotely detectable at ecologically plausible concentrations.

The experiments described above are knowingly simplified since they do not account for the effect of planetary atmosphere on the spectral reflectance, investigating solely the spectrum measured directly above the organisms in sand under controlled laboratory conditions. However, as atmosphere, including clouds, reduces the clarity of spectral features, the results can be considered conservative.

These data do, however, illustrate that even at implausibly high concentrations across a planetary hemisphere, many organisms would exhibit no detectable surface reflectance feature, despite containing pigments with strong absorption features.

Biota that exhibit spectral features with no distinctive absorbances, or absorption features that coincide with abiotic materials [28], will also render a planet devoid of remotely detectable surface biosignatures. A hypothetical example of the latter is the coincidence of the *Deinococcus* green edge (Figure 3b) with the absorption features of many iron oxide compounds such as haematite and goethite [76], which confers upon these compounds a red colouration, similarly to *Deinococcus*. If such organisms were growing on a planet with high concentrations of iron oxides (for example, a planet similar to Mars), even at high concentrations, their absorption edge could be ambiguously mixed with an abiotic green edge.

The simple experiments shown here serve to illustrate the possibility of worlds with surface-dwelling life that is not remotely detectable, based on reflectance spectral features.

**4. Inhabited worlds with biogenic gases indistinguishable from abiotic gases, or at undetectable concentrations**

Surface signatures are not the only means by which astronomers propose to find life on exoplanets. An alternative approach is to search for atmospheric gases at concentrations far above those expected from purely chemical processes or mixtures of gases which would not normally be expected to co-exist. An example of the former signature is a high concentration of oxygen in a planetary atmosphere (at tens of percent concentration), and oxygen mixed with methane is an example of the latter [11]. Oxygen is detectable by proxy via its photochemical product, ozone, formed in the atmosphere from interaction with stellar ultraviolet radiation.

Although these biosignatures have become exemplary and at a distance can be detected in the atmospheric signatures of our present-day planet [15], how might a planet be inhabited, yet exhibit no remotely detectable atmospheric signatures of life? Examples of inhabited worlds with no gaseous signatures of life are discussed in the following section.

## 4.1 Principle of thermodynamic uncertainty

A high atmospheric concentration of oxygen, considered to be one of the most reliable biosignatures, is the product of one particular type of metabolism – oxygenic photosynthesis. This form of metabolism has evolved in only one branch of the terrestrial phylogenetic tree of life (cyanobacteria), being subsequently incorporated into algae and plants by endosymbiotic events [77].

Many gas-producing microbial taxa represented in the Earth's phylogenetic tree of life produce gases that either can also be produced abiotically or are inherited from the protoplanetary disk at high concentrations. Two examples of such metabolisms are shown below.

*Chemolithotrophy*: *Anaerobic iron oxidation*

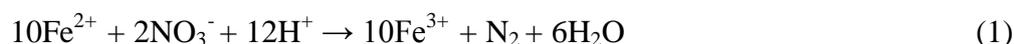
$$10Fe^{2+} + 2NO_3^- + 12H^+ \rightarrow 10Fe^{3+} + N_2 + 6H_2O \qquad (1)$$

*Fermentation*

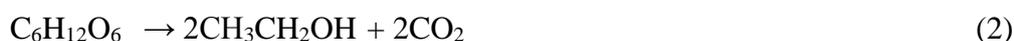
$$C_6H_{12}O_6 \rightarrow 2CH_3CH_2OH + 2CO_2 \qquad (2)$$

In the first reaction (1), anaerobic iron oxidation, reduced iron is used as a source of electrons to reduce nitrate [78-80]. Microbial redox reactions are common in environments where anaerobic conditions prevail and soluble reduced iron (leached from rocks for example) is abundant. This type of metabolism is thought to have been present on the early Earth and is a hypothesised contributor to the production of Banded Iron Formations (BIFs) laid down in the Earth's rock record during the Archaean until ~1.8 Ga ago [81]. However, a product of this reaction is nitrogen gas, present in the Earth's atmosphere at a concentration of 78%. Nitrogen is common in terrestrial-type rocky planetary atmospheres (Mars, 2.7% and Venus, 3.5%).

Another metabolic example that produces a gas common in planetary atmospheres is fermentation [82] (reaction 2), an anaerobic process that involves the substrate-level phosphorylation of organic compounds, with acids, alcohols and carbon dioxide gas products. Carbon dioxide is a common constituent gas in terrestrial-type rocky planets (Mars 95.3%, Venus 96.5%. On the Earth, the concentration is ~0.035%, but in the Earth's early history it would have been much higher [83]).

Nitrogen and carbon dioxide are usually poor biosignature gases on account of at least two characteristics: 1) these gases are produced by abiotic processes or inherited from the protoplanetary disk at high concentrations; and 2) their concentrations in planetary atmospheres are influenced by numerous variables, making it difficult to distinguish putative biotic from abiotic contributions to total atmospheric mixing ratios.

This second characteristic is a particularly important problem. For any exoplanet for which we have an atmospheric spectral signature, we will always lack a great deal of knowledge about geological and geophysical processes, its primordial inventory of gases and the putative influence of a biosphere on atmospheric gas cycles, making it impossible to definitively ascribe a given percentage of gaseous constituents of its atmosphere to biological processes.

Nitrogen gas in the terrestrial atmosphere is an example of this problem. The quantity of nitrogen in biological systems is about three orders of magnitude less than that in the atmosphere, which is about $1.4 \times 10^{20}$ moles [84], making planetary biota a small contributor to atmospheric concentrations. Since there is almost as much nitrogen stored in rocks as in the atmosphere, the question of how atmospheric nitrogen concentrations have varied over time can be reduced to the question of whether rock-atmosphere nitrogen exchange has varied. Berner concludes that because rock-atmosphere nitrogen exchange is small compared to the total mass of nitrogen in the atmosphere, the latter has not varied appreciably, at least over Phanerozoic time [84]. This study highlights at least two important points. Firstly, understanding the factors that control nitrogen concentration in an exoplanet atmosphere requires an understanding of the amount of nitrogen locked in its geosphere and the exchange with the atmosphere. Secondly, if an exoplanet biosphere is comparable to the Earth's, the biota may contribute very minimally to atmospheric nitrogen concentrations, making it impossible to determine whether the atmospheric nitrogen concentration has a biological component.

Microbes not only produce gases, but they also consume them. A reduction in the concentration of certain gas species, relative to levels expected from abiotic processes, might constitute a biosignature. An example of such a gas is carbon monoxide, produced by photolysis of carbon dioxide abiotically, however, also consumed by chemolithotrophic microorganisms [85, 86] (reaction 3).

*Chemolithotrophy*: *Anaerobic carboxydotrophy*

$$CO + NO_3^- \rightarrow CO_2 + NO_2^- \qquad\qquad\qquad (3)$$

Although this approach to life detection seems promising, it also suffers from the problem that trace gases potentially consumed by microorganisms are often part of a cycle of sinks and sources that are poorly understood, even on the Earth. In a discussion about the plausibility of detecting life on Mars by studying whether trace gases have anomalous concentrations caused by biological activity, Weiss et al. [87] concluded that the presence of CO and $H_2$ in the Martian atmosphere suggested a lack of near surface biota to consume them. However, they emphasise our lack of knowledge of putative sources and sinks for these gases, and concluded that the presence of these gases in the atmosphere does not rule out local consumption. These uncertainties highlight another potential problem: a lack of such gases in a planetary atmosphere could also be attributed to unknown abiotic sinks.

All the problems elaborated above may be bundled into a common recognition that we are most likely incapable of determining whether the concentration of numerous exoplanetary atmospheric components deviate thermodynamically from those expected by abiotic processes. We might state this as a general 'problem of exoplanet thermodynamic uncertainty'.

The problem of exoplanet thermodynamic uncertainty can be qualitatively stated as follows: "For many atmospheric gases, the lack of knowledge about an exoplanet, including plate tectonics, hydrosphere-geosphere interactions, crustal geochemical cycling and gaseous sources and sinks, makes it impossible to distinguish a putative biotic contribution to the mixing ratios, regardless of the resolving power of the telescope."

The last part of this problem is important because the lack of knowledge about an exoplanet cannot necessarily be compensated for by improving the quality of the spectrum obtained, if the abiotic concentrations of gases in the atmosphere cannot be accurately modelled. In the context of this paper, the problem raises the possibility of many habitable worlds with no unequivocal remotely detectable atmospheric signs of life.

**4.2 Unequivocal biosignature gases at concentrations below detection**

There are two ways in which a planet with life metabolically capable of producing biosignature gases might remain devoid of biosignatures to alien observers. The metabolising organisms might not have achieved sufficient biomass to change the planetary atmospheric composition. An example being a planet where an origin of life has occurred, biosignature gas-producing organisms have evolved, but they remain at low biomass, perhaps due to competitive disadvantage. A second scenario is a planet where biosignature gas-producing organisms have evolved and achieved high biomass, but the gases subsequently are consumed.

Our own planet, between the Archean and the Great Oxidation Event, is a case in point. Oxygenic photosynthesis is thought to have evolved prior to the rise in atmospheric oxygen to detectable levels (i.e. several percent). Early oxygen is likely to have reacted with reduced compounds, for example reduced iron in the oceans or $H_2S$, $H_2$ and other reduced gas exhalations from volcanism, and been removed from the atmosphere [88].

Reasons for the eventual sudden rise in oxygen in the terrestrial atmosphere remain enigmatic [89, 90]. For this paper, however, the salient point is that a period existed between the evolution of oxygenic photosynthesis (which was at least 2.8 billion years ago, potentially preceding this time [91]) and when gross concentrations of atmospheric oxygen rose ~2.45 billion years ago [83, 88], during which the Earth was devoid of a detectable oxygen biosignature in the atmosphere [92]. This period was at least a tenth of the planetary lifetime, maybe longer. Planets in similar transitions could be habitable, with no remotely detectable unequivocal atmospheric signatures of life.

**4.     Rejecting the hypothesis**

This paper began with the proposed hypothesis: 'Most habitable worlds in the cosmos will have no remotely detectable signs of life'. The preceding sections of this paper gave examples of the multiplicity of ways in which worlds could be habitable and yet have no remotely detectable signatures of life.

Some of these worlds are highly conjectural. Our ability to assess the abundance of Uninhabited Habitable Worlds as a class of exoplanet, for example, is hampered by our lack of knowledge of the frequency and rate of the origin of life on planets that are habitable. If the origin of life is inevitable and rapid on any planet that is habitable, then this class of world may not exist at all. Some of these planets, for example, worlds with biota at high concentrations, but remotely undetectable surface reflectance signatures, can be empirically demonstrated in the laboratory, as shown here, using our own planet's extremophiles.

Nevertheless, we must explore what conditions must be met for the hypothesis proposed here to be rejected. In other words, what general conditions must exist in the Universe for habitable worlds with no remotely detectable signs of life to be numerically in the minority of habitable worlds?

There are four general conditions that must be met in sequence:

1) *Condition 1.* When planets are habitable, the origin of life usually occurs (the fraction of all habitable planets that meet this criterion can be defined as $f_1$).

2) *Conditions 2.* Once life originates, it will usually evolve metabolic processes that produce unequivocal biosignature gases or gaseous disequilibria and/or pigments that have spectral properties easily distinguished from abiotic materials ($f_2$ – the fraction of $f_1$ that meet this criterion).

3) *Condition 3.* Once these organisms evolve, they will usually colonise a planet at high biomass ($f_3$ – the fraction of $f_2$ that meet this criterion).

4) *Condition 4.* Once the organisms colonise the planet, they will usually produce enough gas or surface biosignatures to accumulate at concentrations detectable to alien observers ($f_4$ – the fraction of $f_3$ that meet this criterion).

If we take the word 'most' in the hypothesis to mean 'greater than 50%', then the product of the fraction of planets that meet each of these conditions in sequence (i.e. $f_1$ x $f_2$ x $f_3$ x $f_4$) must be greater than 0.5 for the hypothesis to be rejected. In other words, the rejection of the hypothesis requires that we live in a Universe where the majority of habitable planets have life and sufficient biomass to generate unequivocal biosignatures detectable to us. This would be a remarkable conclusion.

We know that planetary environments for these four conditions to be met can exist. The Earth is an example. However, it is likely that this sequence of events was essential on our own planet for us to be here to write about them. If oxygen produced by oxygenic photosynthesis is required for multicellularity [93], and is required to power large, intelligent brains, then it follows that intelligence is always likely to co-exist on a planet with a remotely detectable biosignature. In other words, there might be an atmospheric biosignature anthropic principle at work.

The hypothesis proposed here, although it focuses on a negative outcome with respect to life detection, is conservative, making the minimum number of assumptions about the likelihood of the origin of life, its abundance and its metabolic trajectories.

In conclusion, the purpose of this paper was to provide astronomers with an experimentally testable, falsifiable biological hypothesis for the study of exoplanets. As the investigation of exoplanets and their spectra expands to greater volumes of the Universe, we will be in a position to accept or reject this hypothesis at different scales.

**Acknowledgements**

I thank the Science and Technology Facilities Council (STFC) for financial support (Consolidated Grant No. ST/1001964/1).

**Figure legends**

**Figure 1.** Four hypothetical scenarios that would result in a class of lifeless exoplanets with habitable conditions (Uninhabited Habitable Worlds).

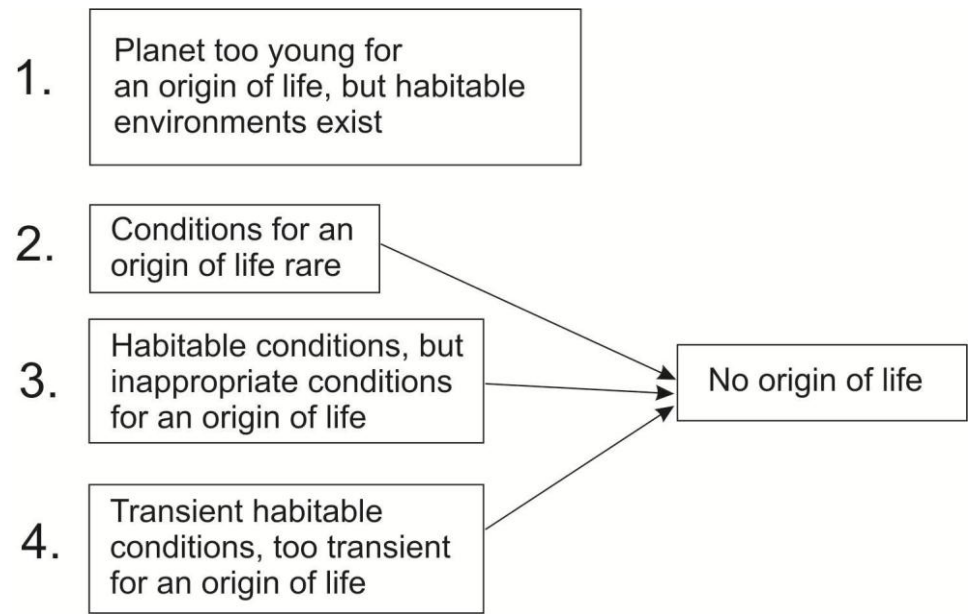

**Figure 2.** Cryptic biota with no surface reflectance spectrum. *Chroococcidiopsis* (A; top image; scale bar 10 μm) and other cyanobacteria inhabit the interstices of a segment of impact-shocked gneiss (B; scale bar 1 cm) from the Haughton impact structure, Canadian High Arctic [55]. The organisms exhibit a typical red-edge from chlorophyll *a* (graph, top curve, vertical dotted line), whilst the surface of the rock (C) exhibits no distinctive spectral feature (graph, bottom curve). The final spectra were the mean of nine separate spectra acquired from three different locations. At each location the spectrum was the mean of ten spectral scans across the wavelength range.

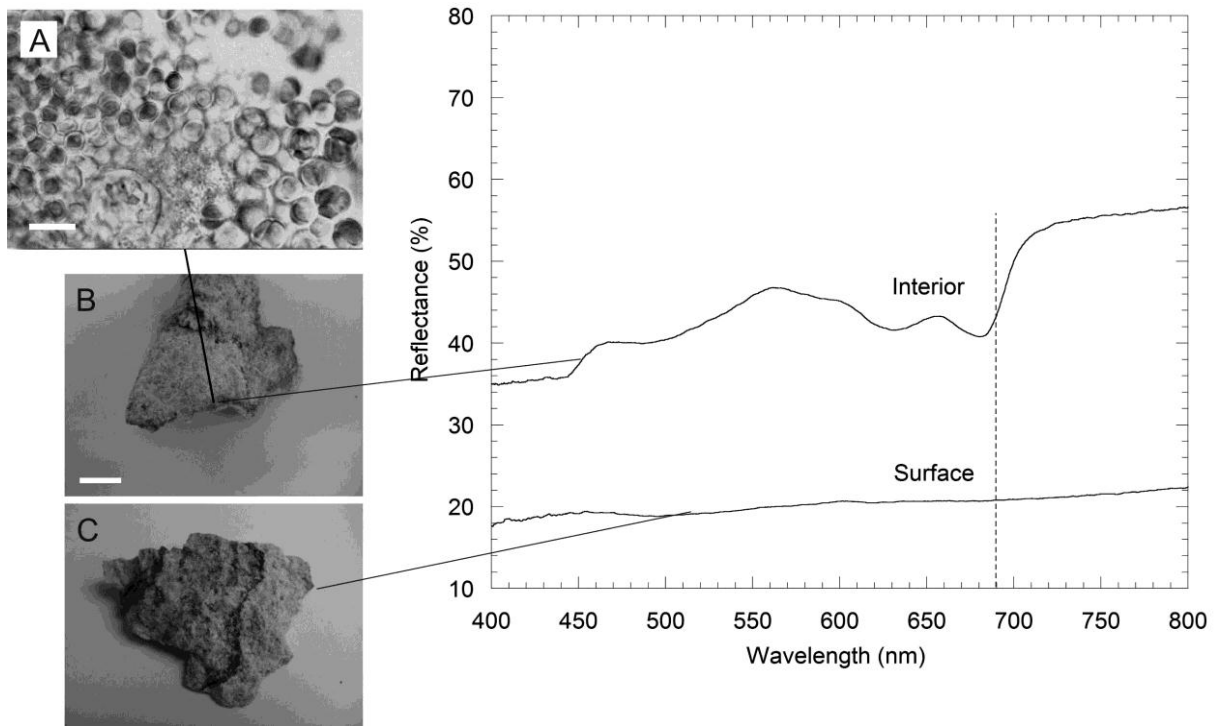

**Figure 3. a.** "*Cupriavidus* world". Spectral absorption features in the microorganism, *C. metallidurans*, are undetectable when mixed with quartz sand at a concentration of $10^8$ organisms/cm$^3$, an implausible concentration across an entire planetary hemisphere. **b.** "*Deinococcus* world". Spectral absorption features in the microorganism, *D. radiodurans*, including its carotenoid-induced 'green edge', are undetectable when mixed with quartz sand at a concentration of $10^8$ organisms/cm$^3$, an implausible concentration across an entire planetary hemisphere. **c.** "*Chroococcidiopsis* world". When mixed with sand at a concentration of $10^6$ organisms/cm$^3$, however, not at 2 x $10^5$ organisms/cm$^3$, the desert cyanobacterium is detectable in quartz sand from its characteristic red edge (vertical dotted line). All spectra are the mean of nine separate spectra acquired from three different locations. At each location the spectrum was the mean of ten spectral scans across the wavelength range. The quantification of the spectral features is provided in Table 1.

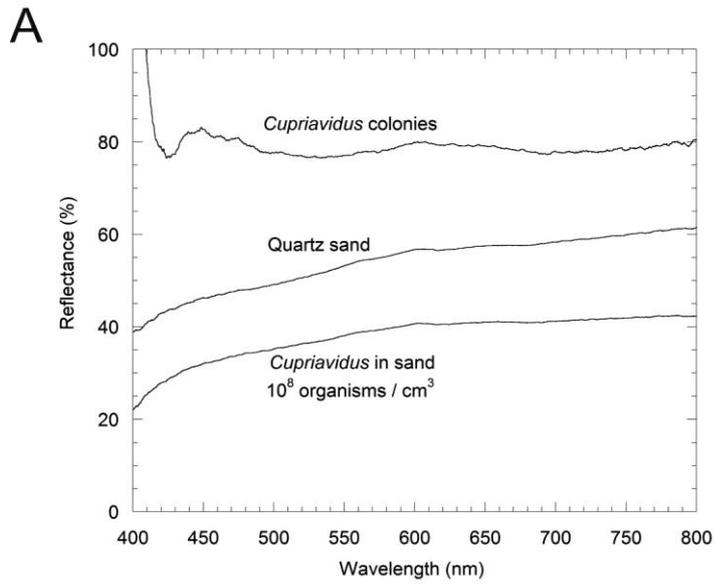

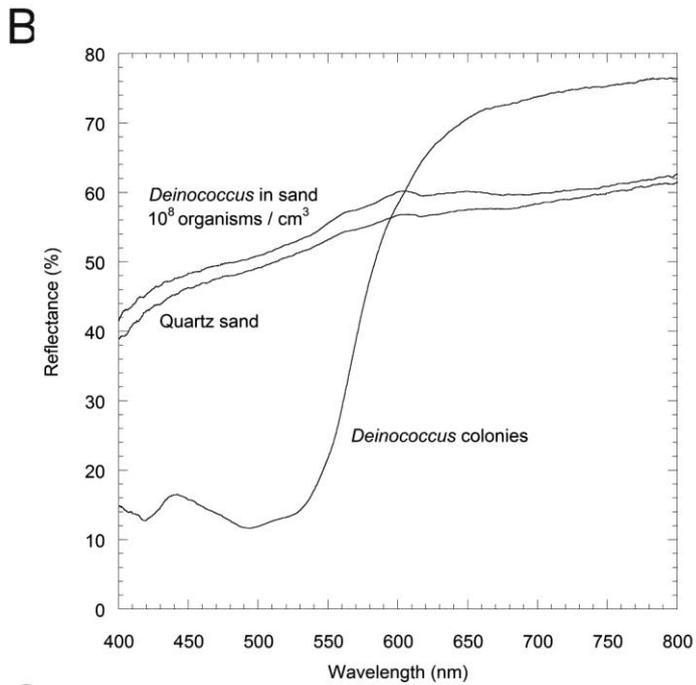

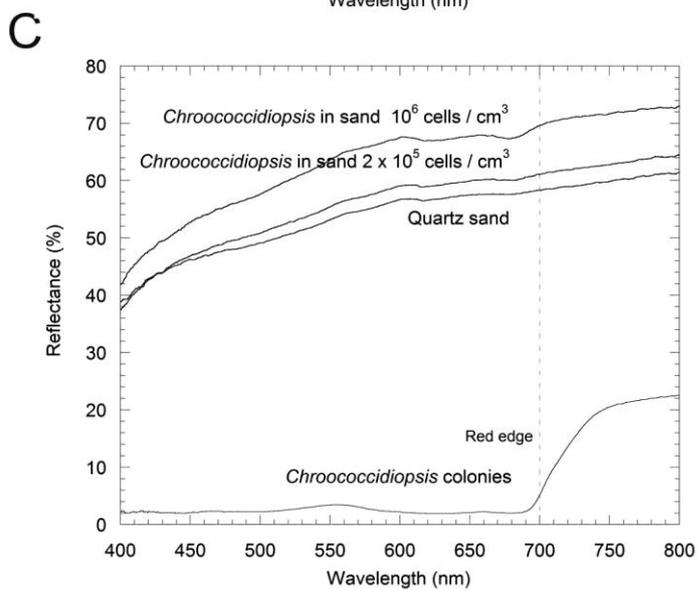

**Table 1.**

The pigment spectral effect (PSE; chlorophyll red edge for *Chroococcidiopsis*) for given organisms mixed with sand, compared to pure colonies and quartz sand controls (Figure 3). PSE was calculated as $(r_b-r_a)/r_a$, where $r_b$ was the mean reflectance from 740-800 nm for all organisms, and $r_a$ was the mean reflectance from 420-480 nm for *C. metallidurans* and *D. radiodurans* and 600-670 nm for *Chroococcidiopsis*. See text for details.

|  | *Cupriavidus metallidurans* | *Deinococcus radiodurans* | *Chroococcidiopsis* |
|---|---|---|---|
| Quartz sand control | 0.33 | 0.32 | 0.06 |
| Bacterial colonies | -0.02 | 4.13 | 21.4 |
| Bacteria and sand | 0.32[a] | 0.28[a] | 0.12[b]; 0.07[c] |

[a] $10^8$ organisms/cm$^3$; [b] $10^6$ organisms/cm$^3$; [c] $2 \times 10^5$ organisms/cm$^3$